# Broadband lightweight flat lenses for longwave-infrared imaging

*Monjurul Meem, [1] Sourangsu Banerji, [1] Apratim Majumder, [1] Fernando Guevara Vasquez, [2] Berardi Sensale-Rodriguez, [1] and Rajesh Menon, [1, 3 a)]*

[1]Department of Electrical and Computer Engineering, University of Utah, Salt Lake City, UT 84112, USA.

[2]Department of Applied Mathematics, University of Utah, Salt Lake City, UT 84112, USA.

[3]Oblate Optics, Inc. 13060 Brixton Place, San Diego CA 92130, USA.

a) rmenon@eng.utah.edu

**ABSTRACT**

We experimentally demonstrate imaging in the longwave-infrared (LWIR) spectral band (8μm to 12μm) using a single polymer flat lens based upon multi-level diffractive optics. The device thickness is only 10μm, and chromatic aberrations are corrected over the entire LWIR band with one surface. Due to the drastic reduction in device thickness, we are able to utilize polymers with absorption in the LWIR, allowing for inexpensive manufacturing via imprint lithography. The weight of our lens is less than 100 times those of comparable refractive lenses. We fabricated and characterized two different flat lenses. Even with about 25% absorption losses, experiments show that our flat polymer lenses obtain good imaging with field of view of ~35° and angular resolution less than 0.013°. The flat lenses were characterized with two different commercial LWIR image sensors. Finally, we show that by using lossless, higher-refractive-index materials like silicon, focusing efficiencies in excess of 70% can be achieved over the entire LWIR band. Our results firmly establish the potential for lightweight, ultra-thin, broadband lenses for high-quality imaging in the LWIR band.

# INTRODUCTION

Longwave infrared (LWIR) imaging refers to imaging in the wavelength band approximately from 8μm to 12μm, and is important for applications ranging from defense [1,2], medicine [3] and agriculture [4] to environmental monitoring [3,5]. In order to attain high transparency, conventional refractive lenses in the LWIR band require materials such as Silicon, Germanium, or Chalcogenide glasses. The weight of these conventional lenses can be too high for many applications. The increased weight limits the range of operation of unmanned aerial vehicles [6]. In addition, such optics render head-mounted night vision goggles heavy, and cause neck and head injuries in soldiers as well as reduce their situational awareness [7]. Here, we show that by appropriately designing thin Multi-level Diffractive Lenses (MDLs), we can correct for image aberrations including chromatic aberrations in the LWIR band and thereby, reduce the weight of such lenses by over 2 orders of magnitude when compared to conventional refractive lenses. In addition, since our MDLs are very thin, i.e. thickness ~ $\lambda_0$, the design wavelength, and the resulting absorption losses are low, we can utilize polymers for the lens material, which are easier to manufacture (for instance, via microimprint lithography).

Conventional refractive optics are comprised of curved surfaces and become thicker with increasing resolution. That is, in order to bend light at larger angles, the radius of curvature must be lowered and consequently, the lens becomes thicker and heavier. Recently, metalenses have been proposed as a means to reduce the thickness of refractive lenses [8-11]. Metalenses are comprised of constituent units that act as antenna elements (of subwavelength thickness), which render a prescribed local phase shift to light upon scattering. By engineering the spatial distribution of such constituent units in the lens plane, it is possible to correct for image aberrations. Although most demonstrations of metalenses have been in the visible and in the near-IR bands, there was a recent example of metalens for one wavelength in the LWIR band, λ=10.6μm [11]. The constituent element of this metalens consisted of a square lattice of cylindrical pillars, whose diameter ranged from 1.5μm to 2.5μm, height = 6.8μm and minimum pitch = 6.2μm. This device demonstrated a focusing efficiency of only 35% at the design wavelength. Another recent demonstration of a metalens-based LWIR microlens also achieved similar performance with similar fabrication challenges [12]. No broadband LWIR metalenses have been demonstrated so far.

We recently showed that when appropriately designed, Multi-level Diffractive Lenses (MDLs) could perform better than metalenses, while being easier to fabricate [13]. Such MDLs have been demonstrated in the THz [14] and in the visible bands [15, 16]. By combining two MDLS, optical zoom has also been demonstrated [17]. In fact, the MDLs require minimum feature width determined approximately by $\min\{\lambda\}/(2*NA)$, where $\min\{\lambda\}$ is the smallest wavelength in the operating spectral band and NA is the numerical aperture of the lens. This feature width is far larger than the corresponding value in the case of metalenses (which tend to be smaller than $\sim\min\{\lambda\}/5$). In addition, MDLs are naturally polarization insensitive and can achieve high efficiencies over large bandwidths and at high NAs [13]. The main drawback of MDLs is their somewhat complex multi-level geometry. However, with modern imprint lithography, such geometries can be manufactured at high volumes and at low costs [18]. Here, we designed several MDLs for the LWIR, fabricated two of them, and then experimentally demonstrated the imaging performance using two different commercially available LWIR image sensors. It is important to distinguish our work from previous reports that utilize Fresnel lenses in the LWIR. An 80μm-thick polymer Fresnel lens combined with a 755μm-thick refractive Silicon lens was used to report the thinnest LWIR lens (total device thickness ~ 0.8mm) demonstrated for imaging [19]. A high-order Fresnel lens made out of Silicon was used in combination with an aperture for wide-angle imaging in the LWIR band as well [20], which had a total device thickness of 1mm. In comparison, the device thickness of our single MDL is only 10µm (a reduction of ~100X) and it is comprised of a patterned polymer. Most importantly, MDLs are corrected for the entire operating bandwidth, while Fresnel lenses are not.

**RESULTS AND DISCUSSION**

First, we designed rotationally symmetric MDLs, whose constituent element is a ring of width equal to 8µm, and whose height is determined by nonlinear optimization. The details of our design methodology are the same as reported in detail before [14-16]. In summary, we maximize the wavelength-averaged focusing efficiency of the MDL, while choosing the distribution of heights of the rings that form the MDL. This optimization is based upon a gradient-descent-assisted direct-binary search. We used an operating band of 8µm to 12µm, and the measured dispersion of a positive-tone photoresist, AZ9260 (Microchem GmbH) in this band (see *Supplementary Information*). We first designed two MDLs with focal length and NA of 19mm and 0.371, and 8mm and 0.45, respectively. Both designs had a constraint of at most 100 height levels. The designed profiles and corresponding simulated point-spread functions (PSFs) are shown in Fig. 1, where close to diffraction-limited focusing at all wavelengths is clearly observed. The full-width at half-maximum (FWHM) of the focal spots were computed for each design wavelength and averaged to obtain a single FWHM to compare to the diffraction-limited FWHM (see *Supplementary Information*). The simulated average FWHM and the diffraction-limited FWHM are 14.3µm and 13.5µm, and 11.2µm and 12.2µm for the MDLs with f=19mm, NA=0.371, and f=8mm, NA=0.45, respectively.

We computed the focusing efficiency of the MDLs as the power within a spot of diameter equal to 3 times the full-width at half-maximum of the spot divided by the total power incident on the lens [**12, 21**]. The focusing-efficiency spectra were computed for all wavelengths of interest and plotted in Fig. S2 of the *Supplementary Information* for the two MDLs shown in Fig. 1. The wavelength averaged (8μm to 12μm) focusing efficiency for the two lenses are 43% and 65%, respectively. The smaller lens has higher efficiency. As described in the *Supplementary Information*, we also computed that about 25% of the incident power is absorbed in the polymer film for both lenses, which accounts for a portion of the reduced focusing efficiency. As described later, it is possible to increase these efficiencies by replacing the polymer with silicon, which is non-absorbing in the LWIR.

We utilized the simulated wavefront after the MDL to compute the equivalent lens aberration. The aberrations are defined as the difference between the simulated wavefront and the ideal spherical wavefront, and the difference is expressed as a linear sum of Zernike polynomials. The coefficients of the Zernike polynomials are illustrated in Fig. 2 for the MDL with NA=0.371, f=19mm computed at $\lambda$=8μm. Similar results were obtained for the other lenses and wavelengths, and included in the *Supplementary Information*. These calculations confirm that MDLs exhibit aberrations that are comparable to or better than those seen in conventional refractive lenses.

The devices were fabricated using grayscale lithography (see *Supplementary Information* for details) [**15-17**]. The optical micrographs of the fabricated lenses are shown in Figs. 3(a) and (b) for the f=19mm and 8mm lenses, respectively. Each lens was then assembled onto a different image sensor; Tau 2 camera core (FLIR) for f=19mm lens (Fig. 3c) and the LW-AAA camera (SeekThermal) for f=8mm lens, whose original lens was manually removed (Fig. 3d). We first characterized the modulation-transfer function (MTF) of the f=19mm, NA=0.371 lens coupled with the Tau 2 sensor [22]. A hot plate with insulator in front was used as an object and the MTF was estimated using the slanted edge (see *Supplementary Information* for details). The temperature of the hot plate was adjusted from $60^0C$ to $140^0C$, and the results are summarized in Fig. 4. There are no significant differences in the MTF with temperature, confirming achromatic imaging.

Several images were taken with both cameras for characterization and these are summarized in Fig. 5. All figures except Fig. 5d are with the f=19mm lens and Tau 2 camera, while Fig. 5d is with the f=8mm lens and LW-AAA (Seek Thermal) camera. Figures 5(a) and 5(d) are of a heated resistor coil, whose diameter is approximately 250μm. The object distance ($z_o$) and the image distance ($z_i$) for each image are labeled in the corresponding figure. By placing a metal block with holes in front of a hot-plate (80°C) at various object distances, we can estimate the resolving power of the camera as indicated in Figs. 5(e)-(l). When the object is 762mm away from the lens, the demagnified image of the holes is spaced by 170μm and these are still well resolved. This spacing corresponds to 10 pixels on the image sensor and represents an angular resolution of ~0.013°. The field of view of the images is about $30^0 X 35^0$ in the horizontal and vertical axes, respectively. Several videos are also obtained from both cameras and have been included as *Supplementary Data*. These include videos of a resistor coil (videos 1 and 2 from the two cameras, respectively), and a human subject indoors (video 3) and outdoors (at night, video 4).

For imaging-efficiency measurements, we used a sharp nail as the object (tip diameter = 4.5mm). The nail was heated to a desired temperature and imaged onto the Tau 2 camera core (FLIR) (see *Supplementary Information* for details and Fig. 6a). The imaging efficiency was estimated as the ratio of the sum of the pixel values inside the spot size to the sum of all the pixel values in the entire frame. The results are summarized in Fig. 6b. An example image at 50°C is shown in Fig 6c. The imaging efficiency was estimated using spot-size of W, 2*W and 3*W as shown in Fig. 6b, where W=0.272mm, the full-width at half-maximum of the demagnified image of the tip of the heated nail. Note that the imaging efficiency is distinct from the focusing efficiency due to the finite size and temperature of the object that is being imaged. In all cases, the efficiency peaks approximately below 60°C. This can be understood by appealing to Wein's law, which determines the peak emission wavelength of a black body at a given temperature (see Fig. 6d). For temperatures above 60°C, the peak wavelength is shorter than 8μm, which is below the designed spectrum of the MDL, and as expected, the efficiency drops. This is further exacerbated by the spectral response of the image sensor, which drops off below ~8μm.

One can utilize higher-refractive index material to increase the focusing efficiency. Since Si exhibits high refractive index and low absorption in the LWIR band (3.42 at $\lambda$=8μm), it is a good candidate material. We designed several MDLs using Si with focal length and NA equal to 19mm and 0.371 respectively. The MDLs were designed with height level constraints of 8, 16, 32 and 64, and the corresponding optimized height profiles are shown in Figs. 7(a)-(d). The corresponding plots of focusing efficiency as function of wavelength are shown in Figs. 7(e)-(h), respectively. Simulated PSFs of all lenses are included in the *Supplementary Information*. With 8 height levels, the Si lens performs approximately equally to the polymer lens with 100 height levels. Once we increase the number of height levels in the Si lens to 16 or higher, the focusing efficiency averaged over all wavelengths is increased significantly to over 71%. Finally, we noticed that the wavelength-averaged efficiency does not increase significantly beyond 16 levels. 16 height levels in Si may be achieved by 4 lithography and etch steps, which are very standard processes in a

CMOS fab [23]. Although this fabrication approach will be more expensive than imprinting directly onto a polymer, in some applications, the additional cost is likely to be justifiable.

**CONCLUSIONS**

Reducing the weight, thickness and number of optical elements will have important applications for all spectral bands. Here, we demonstrate that this can be achieved in the LWIR band using multi-level diffractive lenses. We note that our MDLs are quite distinct from conventional diffractive lenses because of their achromaticity. Conventional diffractive lenses are designed for a specific wavelength and their focusing performance drastically drops at wavelengths away from the design value.

**Methods**

All MDL designs were obtained using nonlinear optimization using a modified gradient-descent-based search algorithm that maximized wavelength-averaged focusing efficiency.

**Acknowledgements**

We thank Brian Baker, Steve Pritchett and Christian Bach for fabrication advice, and Tom Tiwald (Woollam) for measuring dispersion of materials. We would also like to acknowledge the support from Amazon AWS (# 051241749381) for help with the computing facilities. RM and BSR acknowledges funding from the Office of Naval Research grant N66001-10-1-4065 and from an NSF CAREER award: ECCS #1351389, respectively.

**Author Contributions**

RM, BSR and SB conceived and designed the experiments. SB and RM modeled and optimized the devices. MM fabricated the devices. MM and AM performed the experiments and data analysis. FGV performed the MTF analysis. All authors contributed to the paper.

**Competing Interests Statement**

RM is co-founder of Oblate Optics, Inc., which is commercializing technology discussed in this manuscript. The University of Utah has filed for patent protection for technology discussed in this manuscript.

**Materials and Correspondence**

Correspondence and materials requests should be addressed to RM at [rmenon@eng.utah.edu](mailto:rmenon@eng.utah.edu).

**Figures**

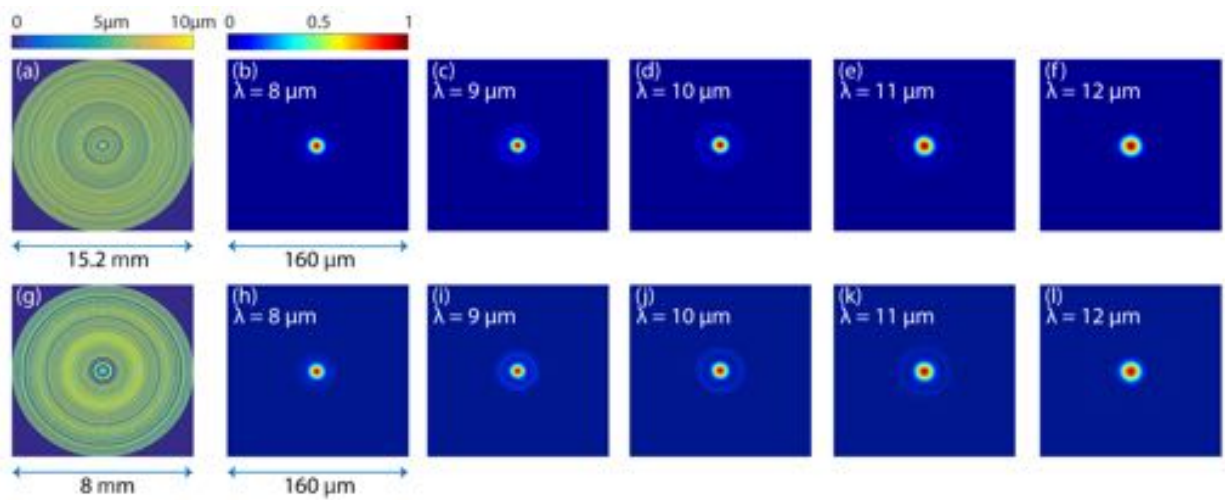

**Fig. 1:** Design and focusing performance of LWIR Multi-level Diffractive Lenses (MDLs). The optimized height profile (a) and (g), and the simulated point-spread functions at the design wavelengths (b-f) and (h-l) for lenses with (focal length, numerical aperture) of top-row (19mm, 0.371) and bottom-row (8mm, 0.45), respectively.

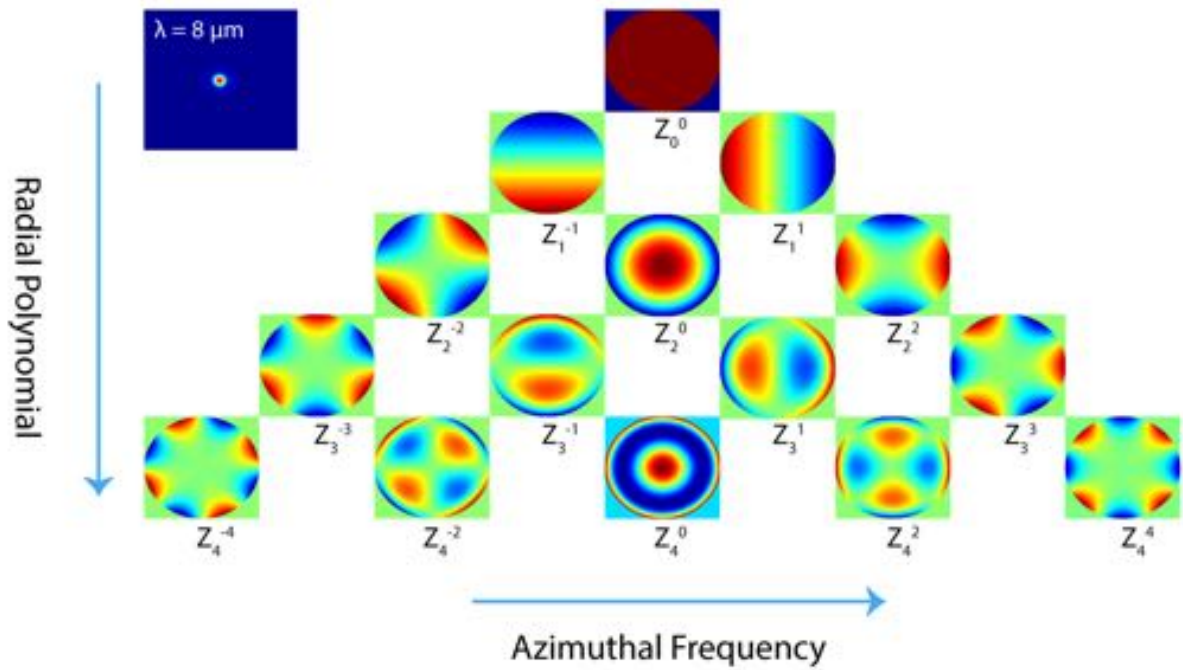

**Fig. 2:** Aberrations analysis of the f=19mm lens (NA=0.371) at λ=8μm. The aberrations coefficients at other design wavelengths are included in the *Supplementary Information*.

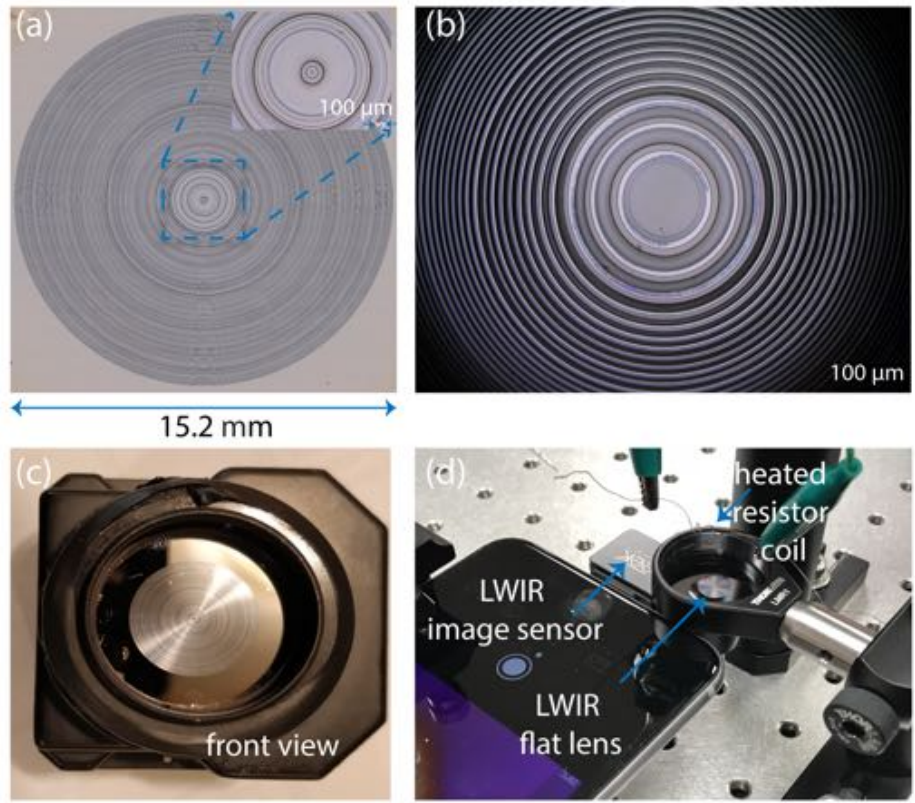

**Fig. 3:** Experiment details. Optical micrographs of the fabricated (a) f=19mm and (b) f=8mm lens. Each lens assembled onto the LWIR image sensor for (c) f=19mm lens with the Tau 2 sensor (FLIR) and (d) f=8mm lens with the LW-AAA sensor (SeekThermal).

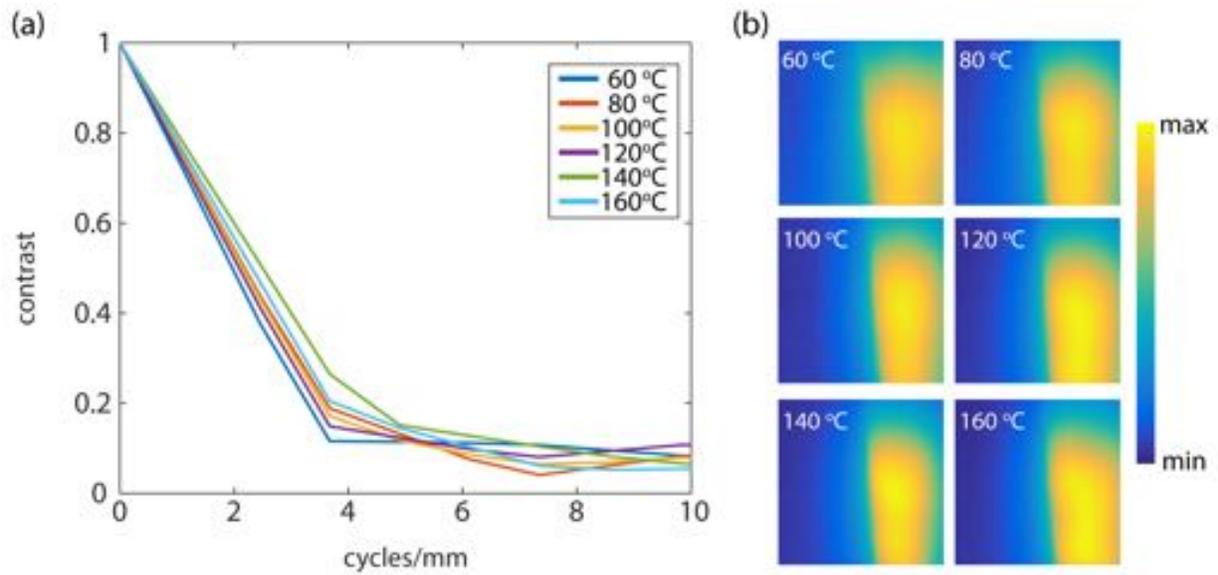

**Fig. 4**: Modulation-transfer function (MTF) of lens with f=19mm, NA=0.371 and Tau 2 image sensor (FLIR). (a) MTF curves for different temperatures show good consistency. (b) Raw images used to compute the MTF curves.

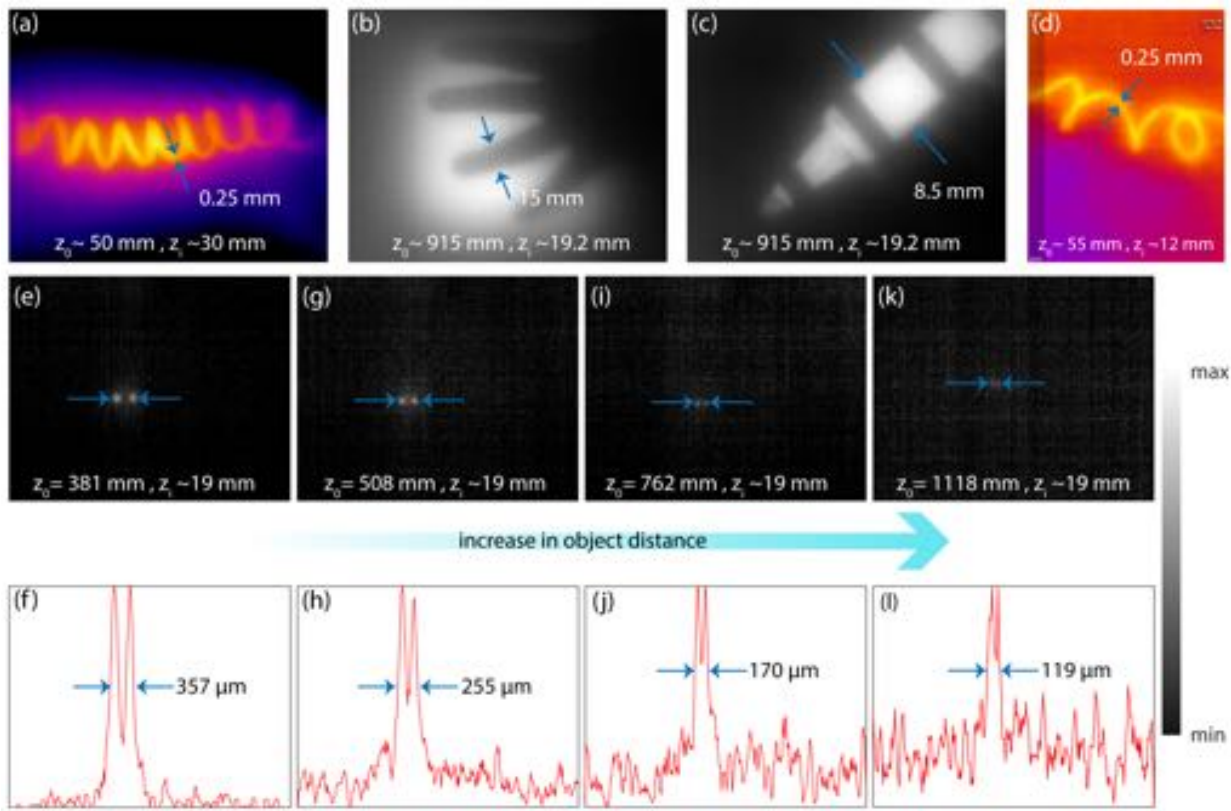

**Fig. 5**: Exemplary images taken using the flat-lens LWIR cameras. All images except (d) are taken with the f=19mm lens and Tau 2 core, (d) is with f=8mm and the LW-AAA (SeekThermal) camera. Object distance = $z_0$ and image distance = $z_i$ are labeled in the figures. The images (e,g,i,k) and corresponding linescans (f, h, j, l) of two holes in a metal block placed in front of a hot-plate heated to 80°C taken at increasing distances from the camera. The holes are well resolved at a distance as large as 762mm, which corresponds to an angular resolution of ~0.013⁰.

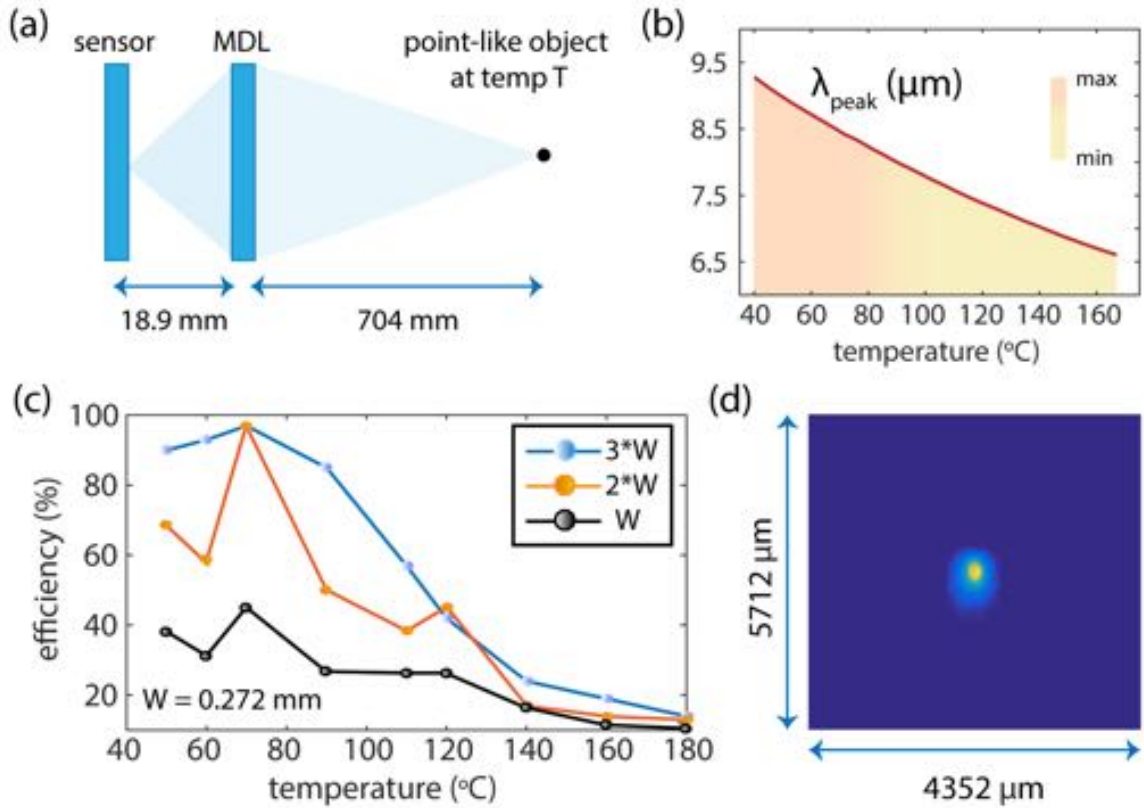

**Fig. 6**: Characterization of focusing efficiency. (a) Schematic of experiment. (b) Peak wavelength corresponding to a blackbody temperature using Wein's law, showing that efficiency peak occurs for temperatures of ~50°C, which corresponds to $\lambda_{peak}$~8.5µm. (c) Focusing efficiency of f=19mm lens with Tau 2 camera core as a function of the object (hot plate) temperature. (d) Exemplary point-spread function at 50°C (raw data).

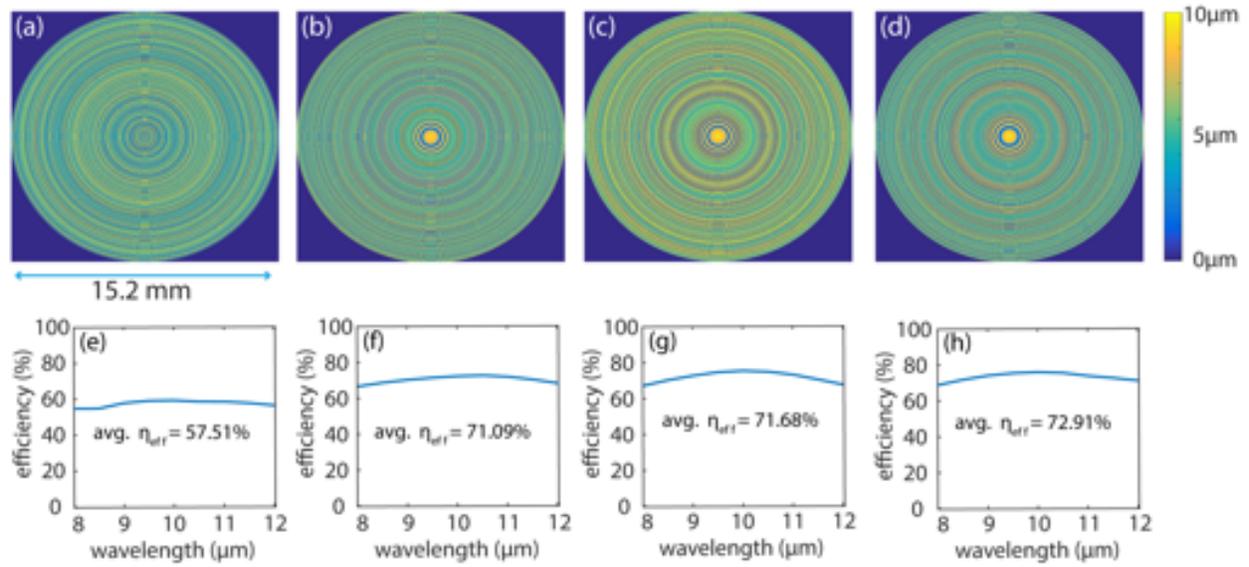

**Fig. 7**: The optimized height profile of Si MDL for NA=0.371 and focal length=19mm with (a) 8 levels (b) 16 levels (c) 32 levels and (d) 64 levels. The corresponding simulated focusing efficiencies for the MDL with (e) 8 levels (f) 16 levels (g) 32 levels and (h) 64 levels. It is observed that the focusing efficiency tends to improve only marginally beyond 16 levels in this case. All other design parameters are the same.

# Supplementary Information

# Broadband lightweight flat lenses for longwave-infrared cameras

**Monjurul Meem, [1] Sourangsu Banerji, [1] Apratim Majumder, [1] Fernando Guevara-Vasquez, [2] Berardi Sensale-Rodriguez[1] & Rajesh Menon[1, 3]***

[1]*Dept. of Electrical & Computer Engineering, University of Utah, 50 Central Campus Dr. Salt Lake City UT 84112 USA*
[2]*Dept. of Applied Mathematics, University of Utah, 50 Central Campus Dr. Salt Lake City UT 84112 USA.*
[3]Oblate Optics, Inc. 13060 Brixton Place, San Diego CA 92130 USA.
*\* rmenon@eng.utah.edu*

## 1. Dispersion of AZ9260 and Si in LWIR band

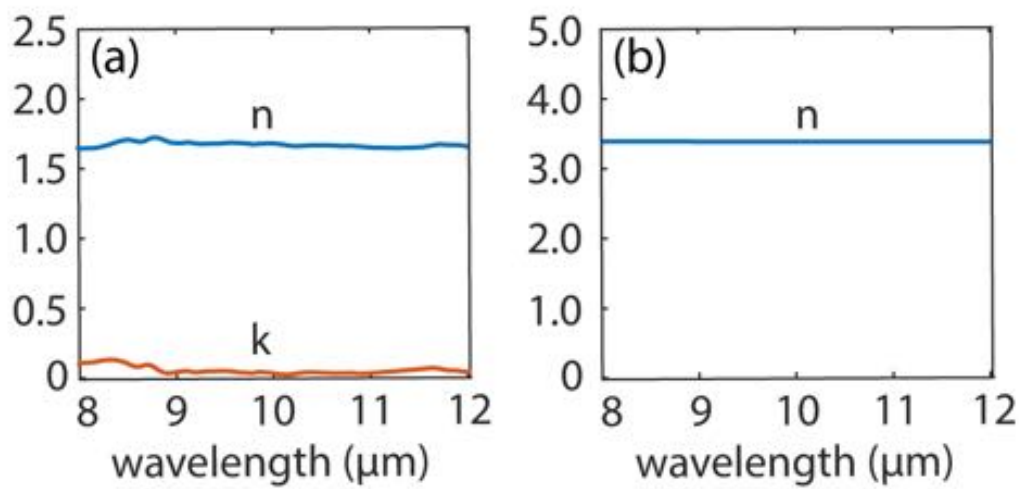

***Fig. S1:*** *Dispersion of (a) AZ9260 and (b) Silicon in the LWIR band.*

## 2. Focusing efficiency spectra of the 2 lenses in Fig. 1.

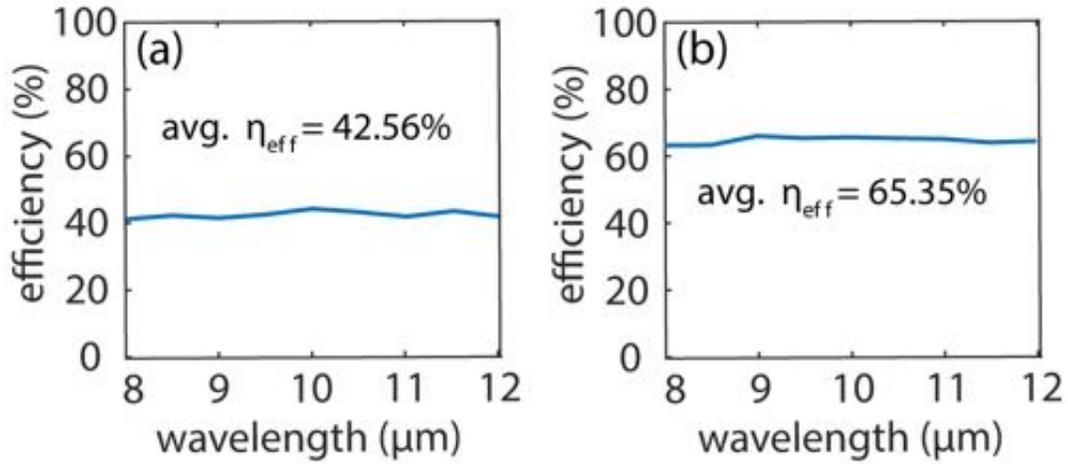

***Fig. S2**: Focusing efficiency spectra of lens with (a) f=19mm, NA=0.371 and (b) f=8mm, NA=0.45. Both designs are shown in Fig. 1 of the main text.*

## 3. Full-width at half-maximum of the focal spots of the 2 lenses in Fig. 1.

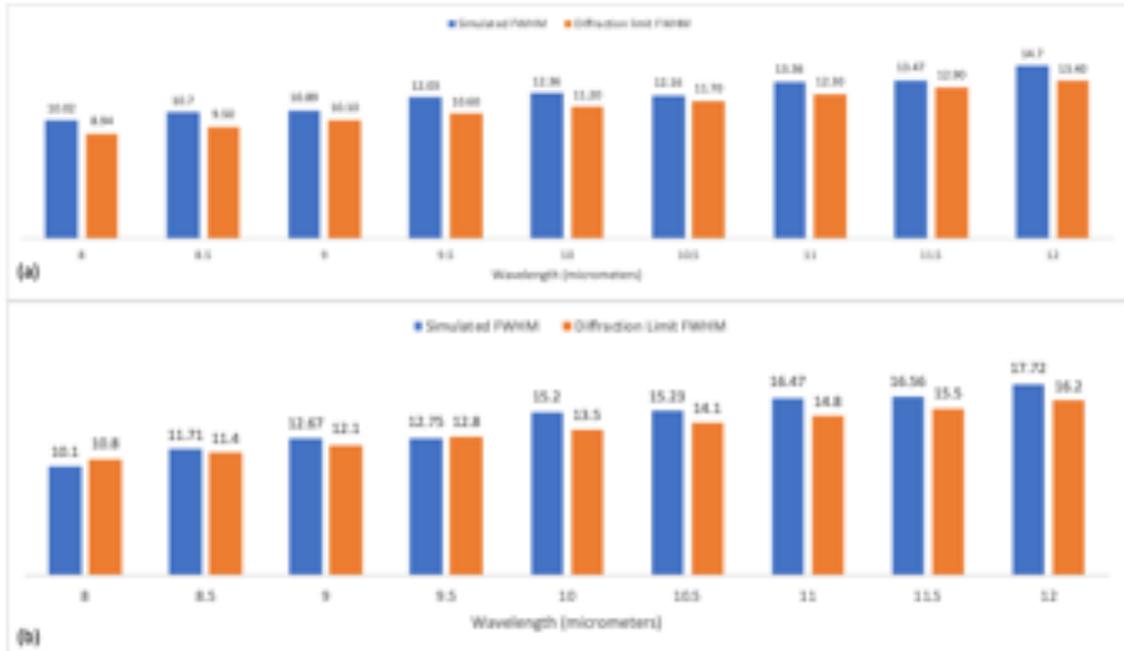

***Fig. S3:** Simulated full-width at half-maximum (FWHM) of the focal spots (shown in blue) of lens with (a) f=19mm, NA=0.371 and (b) f=8mm, NA=0.45. The corresponding diffraction limited FWHM are shown in orange for comparison. Both designs are shown in Fig. 1 of the main text.*

## 4. Absorption in polymer film of the 2 lenses in Fig. 1.

The absorption in the film was computed using the extinction coefficient, $\alpha = 4\pi k/\lambda$, where k is the complex part of the measured refractive index. The absorbed fraction is then computed as $1 - \exp(-\alpha*h)$, where h is the ring height of each ring in the lens. We can then average the absorption across all the rings to compute an estimate of the total absorption.

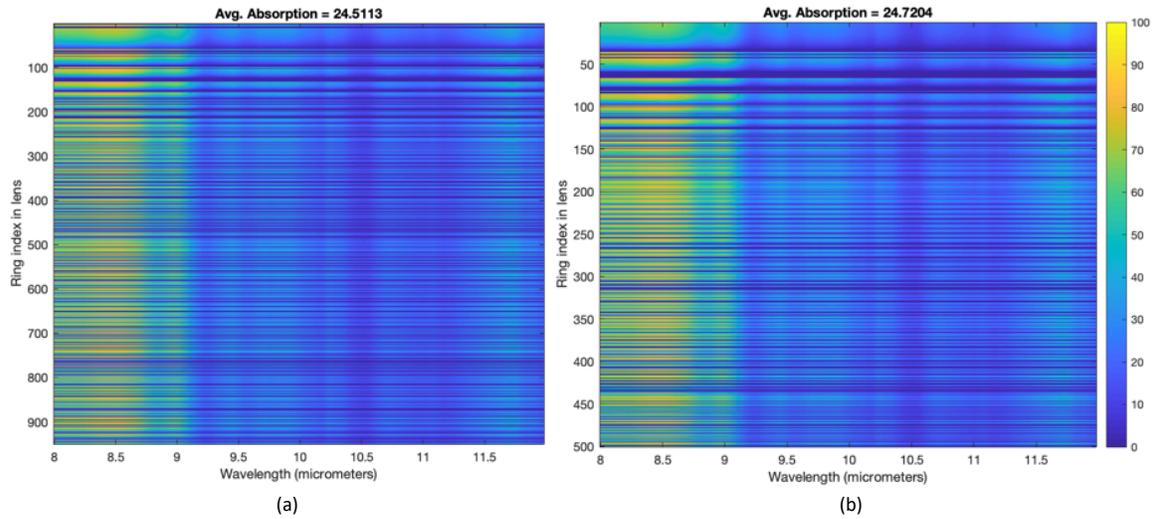

*Fig. S4:* Calculated absorption in the AZ9260 film as function of wavelength and ring index for lens with (a) f=19mm, NA=0.371 and (b) f=8mm, NA=0.45. The absorption averaged over the entire lens is shown on top for each lens. Both designs are shown in Fig. 1 of the main text.

## 5. Aberrations Analysis

The Zernike polynomial coefficient were fitted over a circular shaped pupil. The calculation was done using the reference [**57**]. The fit was using the least squares fit method. Fringe indexing scheme was used.

**Table S1: Aberrations coefficients**

| Radial degree (n) | Azimuthal degree (m) | Fringe index (j) | Classical name |
|---|---|---|---|
| 0 | 0 | 1 | piston |
| 1 | 1 | 2 | tip |
| 1 | -1 | 3 | tilt |
| 2 | 0 | 4 | defocus |

| 2 | 2 | 5 | vertical astigmatism |
| 2 | -2 | 6 | oblique astigmatism |
| 3 | 1 | 7 | horizontal coma |
| 3 | -1 | 8 | vertical coma |
| 4 | 0 | 9 | primary spherical |
| 3 | 3 | 10 | oblique trefoil |
| 3 | -3 | 11 | vertical trefoil |
| 4 | 2 | 12 | vertical secondary astigmatism |
| 4 | -2 | 13 | oblique secondary astigmatism |
| 4 | 4 | 14 | vertical quadrafoil |
| 4 | -4 | 15 | oblique quadrafoil |

The following lists all the fitting coefficients for the designed lenses:

**Table S2: Flat Lens with NA = 0.371, f = 19 mm**

| Wavelength (um) | Piston | Tip | Tilt | Defocus | Vertical astigmatism | Oblique astigmatism | Horizontal coma | Vertical coma | Primary spherical | Oblique trefoil | Vertical trefoil | Vertical secondary astigmatism | Oblique secondary astigmatism | Vertical quadrafoil | Oblique quadrafoil |
|---|---|---|---|---|---|---|---|---|---|---|---|---|---|---|---|
| 8 | 2.42E-05 | -1.14E-08 | -1.14E-08 | -4.50E-05 | -2.01E-21 | 1.13E-11 | 2.49E-08 | 2.49E-08 | 3.50E-05 | 1.09E-09 | 1.09E-09 | -1.55E-20 | 3.41E-11 | -6.17E-06 | -5.46E-23 |
| 8.5 | 2.04E-05 | -9.01E-09 | -9.01E-09 | -3.79E-05 | 1.93E-21 | -2.34E-11 | 1.98E-08 | 1.98E-08 | 3.09E-05 | 1.14E-09 | 1.14E-09 | -6.88E-22 | -2.79E-11 | -5.14E-06 | 7.36E-23 |
| 9 | 2.02E-05 | -4.36E-09 | -4.36E-09 | -3.98E-05 | -7.17E-22 | -1.55E-11 | 5.80E-09 | 5.80E-09 | 3.93E-05 | 6.73E-10 | 6.73E-10 | -2.55E-20 | -2.58E-11 | -3.95E-06 | 1.15E-23 |
| 9.5 | 2.23E-05 | -5.33E-09 | -5.33E-09 | -4.30E-05 | -3.01E-21 | 2.56E-11 | 7.94E-09 | 7.94E-09 | 4.14E-05 | 1.50E-09 | 1.50E-09 | -1.67E-21 | 4.28E-11 | -4.51E-06 | 4.55E-23 |
| 10 | 1.89E-05 | -4.99E-09 | -4.99E-09 | -3.63E-05 | 1.07E-20 | -1.83E-11 | 3.19E-09 | 3.19E-09 | 3.56E-05 | -1.33E-09 | 1.33E-09 | -1.80E-21 | -3.04E-11 | -3.57E-06 | -8.18E-23 |
| 10.5 | 2.01E-05 | -3.98E-09 | -3.98E-09 | -3.78E-05 | 7.35E-22 | -1.41E-11 | 5.42E-09 | 5.42E-09 | 3.61E-05 | 3.43E-11 | -3.43E-11 | -6.83E-21 | -2.41E-11 | -3.82E-06 | -1.95E-23 |
| 11 | 2.57E-05 | -5.09E-09 | -5.09E-09 | -4.76E-05 | 5.49E-21 | 1.34E-10 | 6.59E-09 | 6.59E-09 | 4.38E-05 | 8.18E-10 | 8.18E-10 | -6.53E-21 | 2.22E-10 | -4.82E-06 | -2.14E-23 |
| 11.5 | 2.41E-05 | -5.53E-09 | -5.53E-09 | -4.32E-05 | -1.12E-20 | 1.29E-11 | 1.18E-08 | 1.18E-08 | 3.67E-05 | 1.97E-09 | 1.97E-09 | -1.68E-20 | 2.39E-11 | -5.04E-06 | -1.03E-22 |
| 12 | 1.77E-05 | -3.62E-09 | -3.62E-09 | -3.17E-05 | 2.54E-21 | -7.91E-12 | 7.00E-09 | 7.00E-09 | 2.72E-05 | 1.13E-09 | 1.13E-09 | 7.39E-21 | -1.17E-11 | -3.54E-06 | -2.11E-23 |

**Table S3: Flat Lens with NA = 0.45, f = 8 mm**

| Wavele | Pisto | Tip | Tilt | Defoc | Vertical | Oblique | Horizo | Verti | Prima | Obliq | Verti | Vertical | Oblique | Vertica | Obliqu |

| ngth (um) | n | | | us | astigmatism | astigmatism | ntal coma | cal coma | ry spherical | ue trefoil | cal trefoil | secondary astigmatism | secondary astigmatism | l quadrafoil | e quadrafoil |
|---|---|---|---|---|---|---|---|---|---|---|---|---|---|---|---|
| 8 | 1.08E-05 | -5.77E-09 | -5.77E-09 | -2.66E-05 | 2.71E-21 | -3.38E-12 | 1.70E-08 | 1.70E-08 | 3.10E-05 | -7.64E-10 | 7.64E-10 | -2.00E-21 | 7.98E-12 | -3.74E-07 | -5.61E-24 |
| 8.5 | 9.09E-06 | -4.71E-09 | -4.71E-09 | -2.22E-05 | -2.62E-21 | -3.54E-12 | 1.36E-08 | 1.36E-08 | 2.61E-05 | -7.66E-10 | 7.66E-10 | 1.87E-21 | 5.07E-12 | -2.68E-07 | 9.79E-24 |
| 9 | 9.00E-06 | -2.23E-09 | -2.23E-09 | -2.23E-05 | -8.18E-22 | -1.47E-12 | 6.37E-09 | 6.37E-09 | 2.77E-05 | -3.68E-10 | 3.68E-10 | -5.04E-22 | 1.98E-12 | -1.46E-07 | 2.04E-24 |
| 9.5 | 1.00E-05 | -2.87E-09 | -2.87E-09 | -2.47E-05 | 1.10E-21 | 5.52E-13 | 7.18E-09 | 7.18E-09 | 3.02E-05 | -7.48E-10 | 7.48E-10 | -1.45E-22 | 5.82E-12 | -2.08E-07 | -4.30E-25 |
| 10 | 8.53E-06 | -1.72E-09 | -1.72E-09 | -2.09E-05 | -5.65E-22 | -3.43E-12 | 4.83E-09 | 4.83E-09 | 2.55E-05 | -3.08E-10 | 3.08E-10 | 6.60E-21 | -3.12E-12 | -2.20E-07 | -1.41E-23 |
| 10.5 | 9.06E-06 | -1.53E-09 | -1.53E-09 | -2.19E-05 | 1.41E-23 | 2.39E-12 | 5.02E-09 | 5.02E-09 | 2.65E-05 | -5.63E-11 | 5.63E-11 | 7.55E-21 | 6.04E-12 | -2.88E-07 | 3.20E-24 |
| 11 | 1.16E-05 | -1.71E-09 | -1.71E-09 | -2.78E-05 | 2.06E-21 | 2.74E-12 | 6.24E-09 | 6.24E-09 | 3.32E-05 | 1.78E-10 | 1.78E-10 | -3.43E-21 | 6.53E-12 | -3.75E-07 | -2.53E-24 |
| 11.5 | 1.08E-05 | -2.15E-09 | -2.15E-09 | -2.56E-05 | -1.23E-21 | 6.82E-13 | 8.51E-09 | 8.51E-09 | 2.99E-05 | 2.79E-10 | 2.79E-10 | -5.00E-21 | 5.27E-12 | -3.98E-07 | -9.79E-24 |
| 12 | 7.97E-06 | -1.09E-09 | -1.09E-09 | -1.88E-05 | -3.36E-21 | 1.82E-12 | 4.46E-09 | 4.46E-09 | 2.20E-05 | 1.77E-10 | 1.77E-10 | -4.37E-22 | 4.02E-12 | -2.63E-07 | -7.94E-25 |

## 6. Fabrication

The MDLs were fabricated using direct laser write grayscale lithography [1]. Previously we have demonstrated grayscale lithography with positive thin photoresist [2], this time used a thick photoresist, as we needed taller pixels for LWIR lens. A positive tone photoresist (AZ9260) [3] was spin coated on a 1" double sided polished Si wafer at 2000 rpm for 60 seconds to yield a thickness of 10μm. The spin-coated sample was baked in an oven at 110°C for 30 minutes. The samples were then left overnight inside a rehydration chamber (RH~60%). The designs were written on the sample using a Heidelberg μPG 101 [4] tool and developed in AZ 300 MIF developer [5] solution for 15 minutes. A calibration step was performed beforehand to determine the exposed depths at a particular gray scale level. The nonlinear relationship between the exposed depth and grayscale level depends on a given set of process variables - i.e. photoresist type, photoresist thickness, hard/soft baking time, humidity, developer used, developing time, Heidelberg µPF 101 settings and so on. A new calibration has to be done whenever there is a change in any aspect of the photoresist processing. The tool we used allows 100 different grayscale levels. The exposed depth profile vs grayscale level for a particular calibration is given fig. S5.

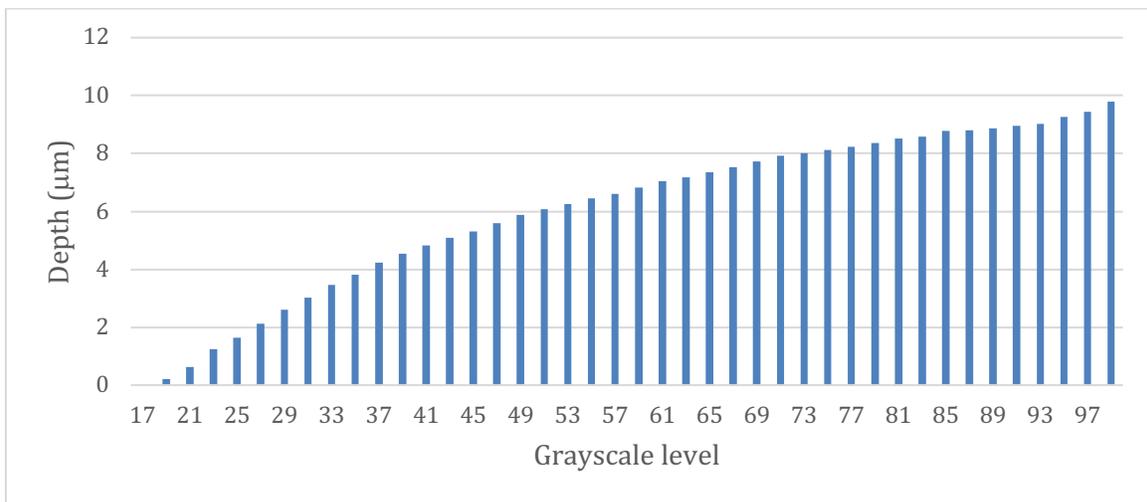

**Fig. S5:** *Exposed depth vs grayscale level*

## 7. Metrology of the fabricated lens

We measured the heights of the fabricated design (f=19mm, NA=0.371) over 10 randomly selected rings using a stylus profilometer (Tencor P-10). The results are summarized in fig. S6. The estimated error has a mean of 1.28µm and standard deviation of 800nm.

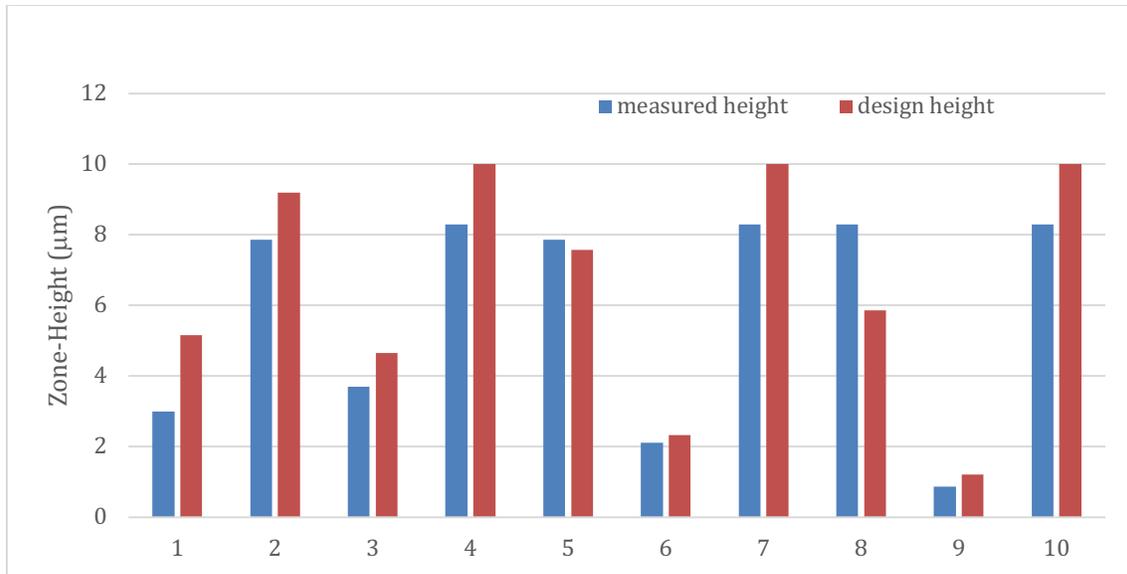

*Fig. S6: Ring-height measurement.*

## 8. Experiment Details

For imaging experiments, the LWIR lens was mounted on a xyz translation stage. Different hot objects were imaged onto a thermal sensor. We used two different thermal sensors; Tau 2 336 thermal core (FLIR) and LW-AAA thermal sensor (SeekThermal). The experimental setup is demonstrated in fig. S7.

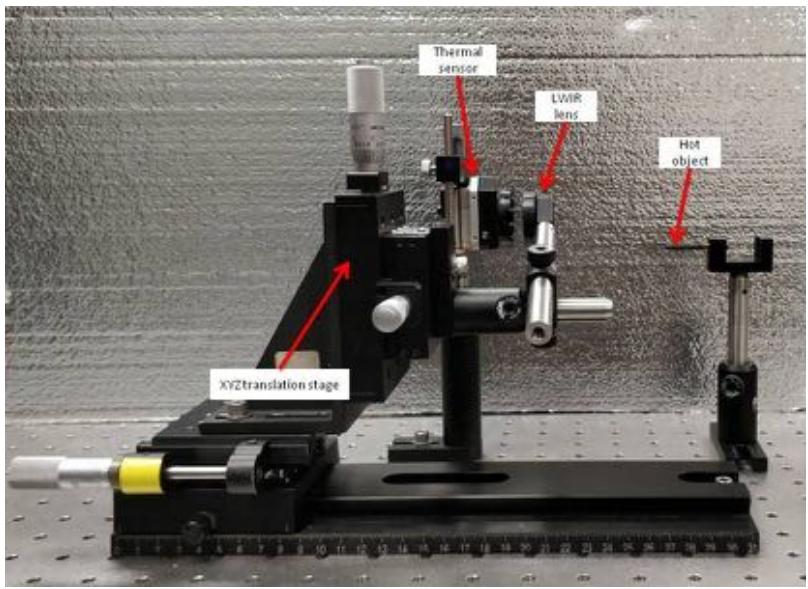

*Fig. S7: Experimental setup for imaging.*

For Imaging, different items were used as object and imaged onto the thermal sensor using LWIR lens. Some of the objects are presented in fig. S8.

For MTF calculation we imaged the slanted edge of an insulator (Plexiglas + Styrofoam) with a hotplate (Model: HS 30, Torrey Pines Scientific) behind it, as shown in fig. S8 (b). The temperature of the hotplate was varied from 60°C to 160°C. For all slanted edge images at different temperature, the object distance and image distance were kept fixed at 508mm and 20mm respectively. In each case, an image of the background with no heat source in front was taken, which served as dark frame and was subtracted from the images.

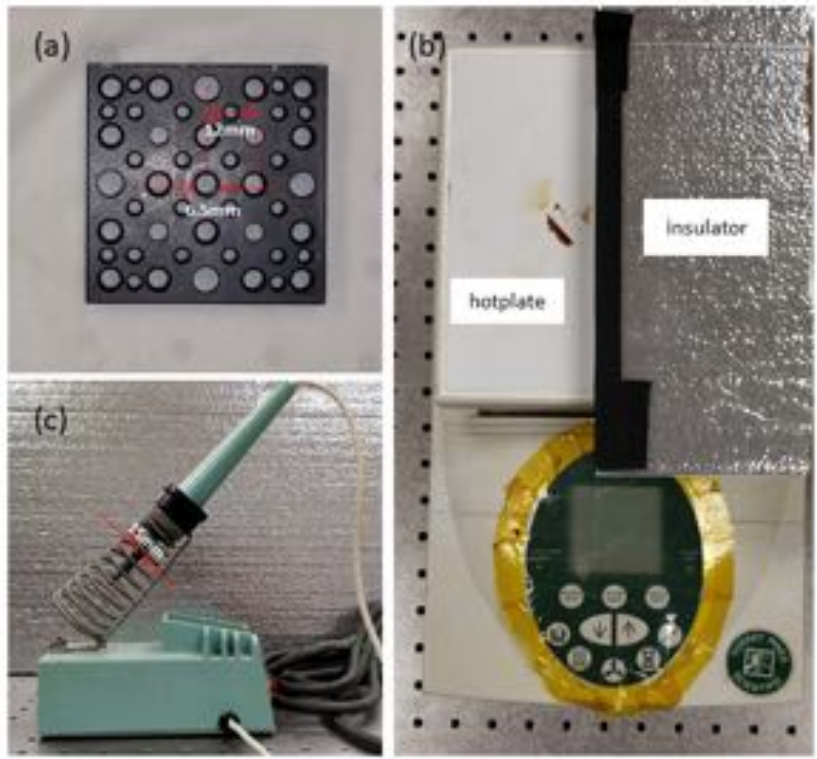

*Fig. S8:* *Objects used for imaging. (a) metal block with holes, used in main fig.5 (e) - (i), (b) hotplate with insulator in front, used for MTF characterization, (c) soldering iron, used for fig. 5(c)*

For efficiency measurement, we imaged a nail (diameter = 4.5mm) as an object (shown in fig. S8). The nail was heated to the desired temperature with a torch and then was imaged onto the Tau2 thermal sensor. A dark frame was also captured using aforementioned technique and subtracted from the images as well. The efficiency was estimated using the formula, efficiency = sum of pixel values in 3*W / sum of pixel values in full frame. The PSFs and W used for the efficiency measurements are given in fig. S9 and fig. S10 respectively.

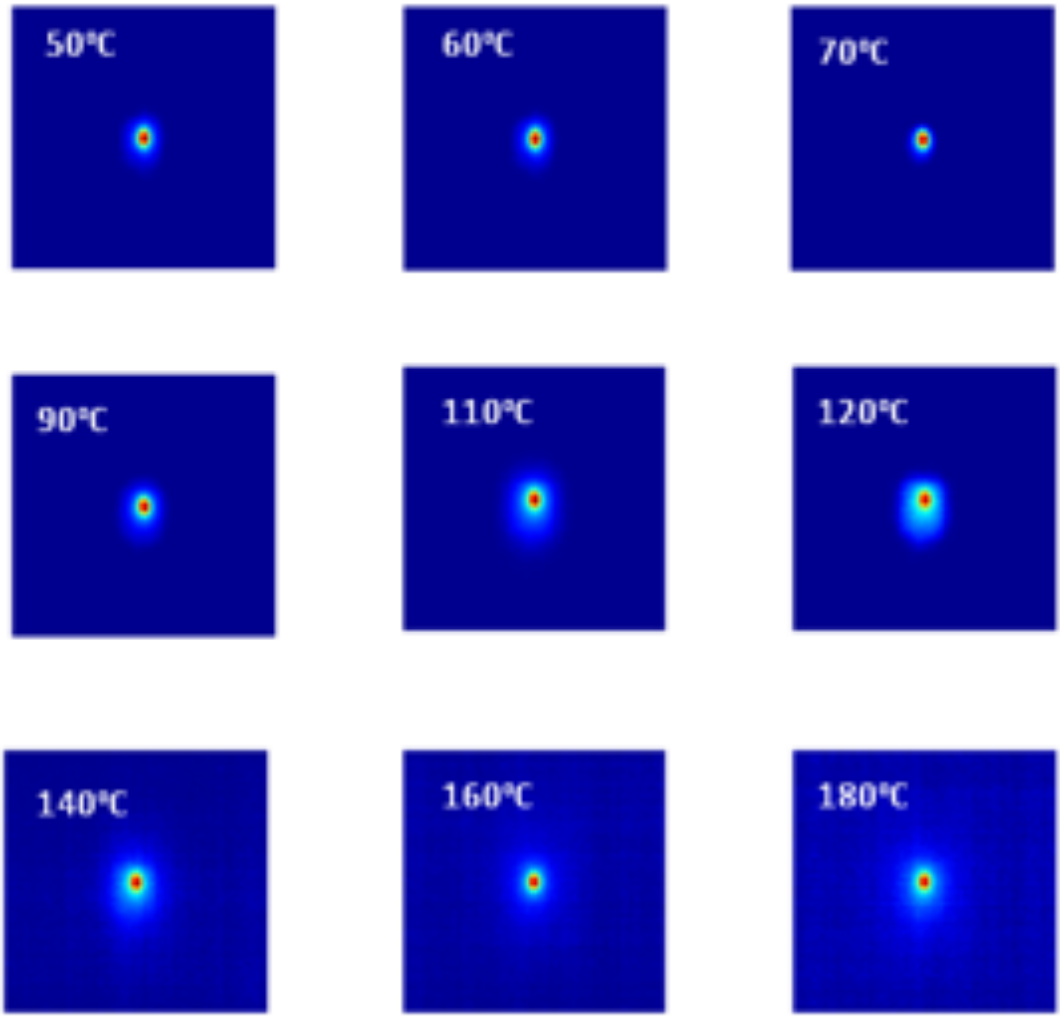

***Fig. S9:*** *PSF's at different temperatures*

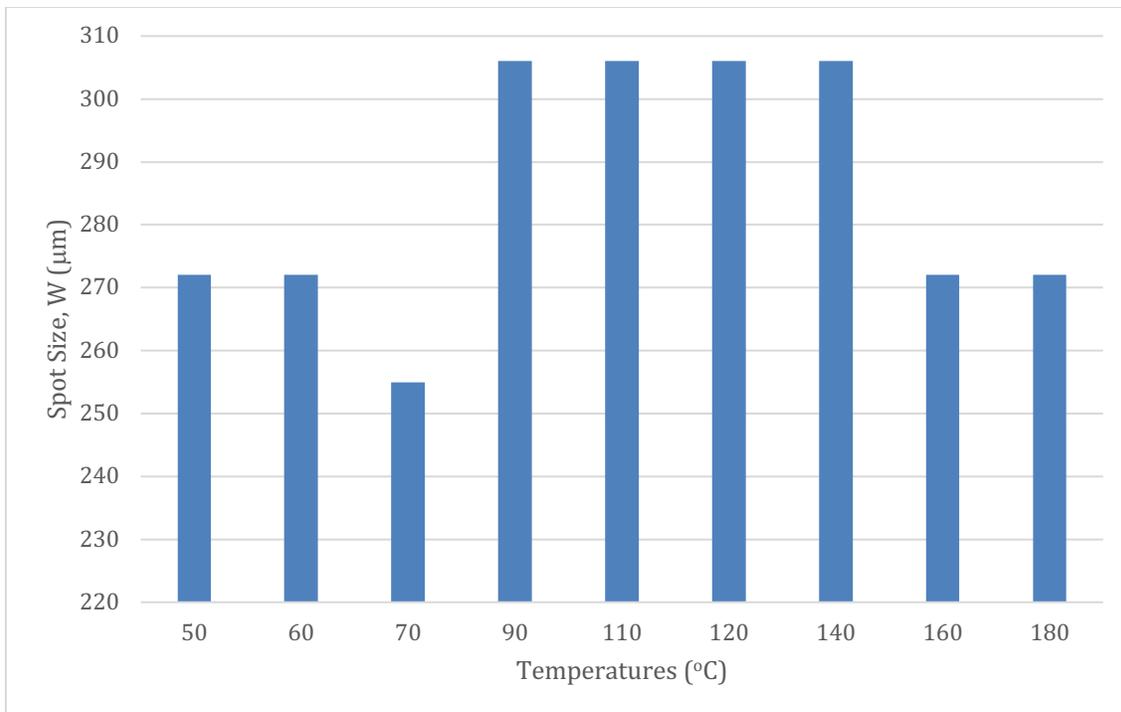

***Fig. S10:*** *Spot size (W) at different temperatures.*

## 9. Camera Assembly:

We put the LWIR lens on a 3D printed lens mount and attached it directly to Tau2 thermal sensor to make a standalone thermal camera. The demo camera is illustrated in fig S11. The lens mount can be adjusted for better focusing as well.

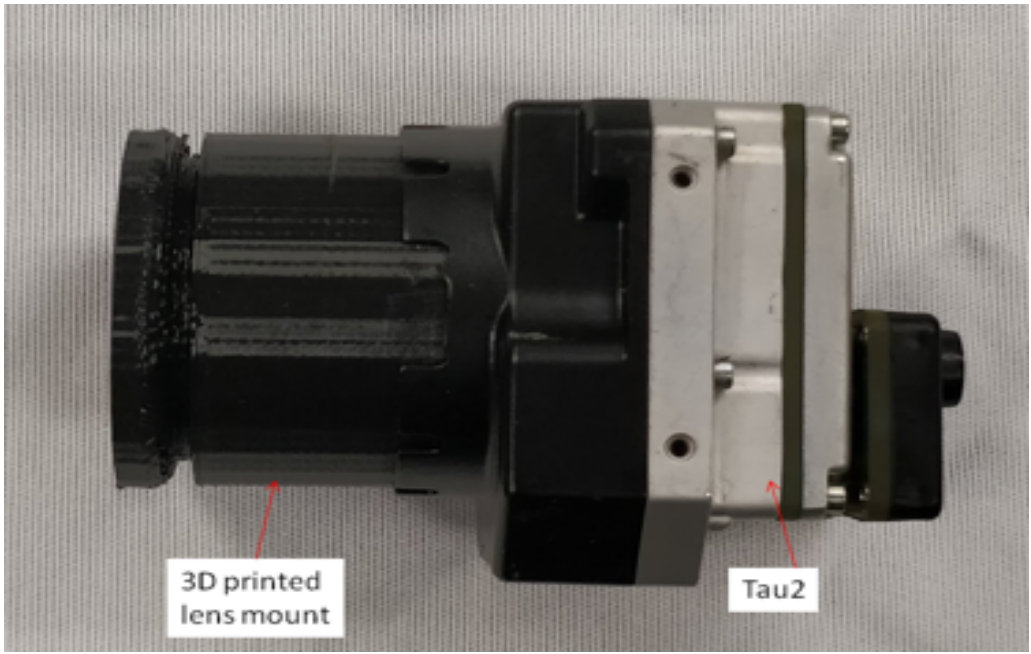

*Fig. S11:* Demo camera assembly

## 10. Simulated Point Spread Functions (PSFs) [Si Flat Lenses]

The simulated point spread functions (PSFs) for the broadband lenses designed with 8, 16, 32 and 64 height-quantized levels are provided below:

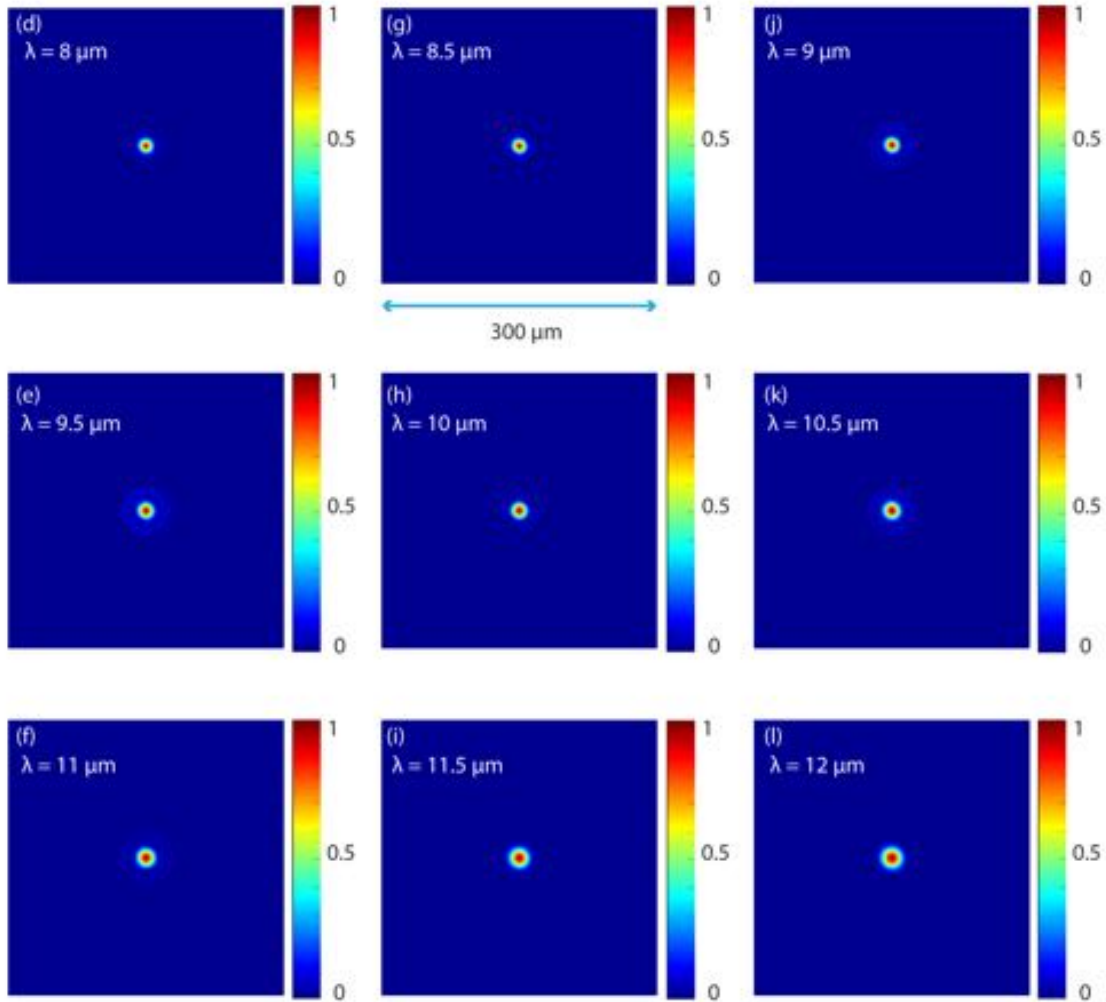

*Fig. S12:* *Simulated PSFs of the 8-level Si lens.*

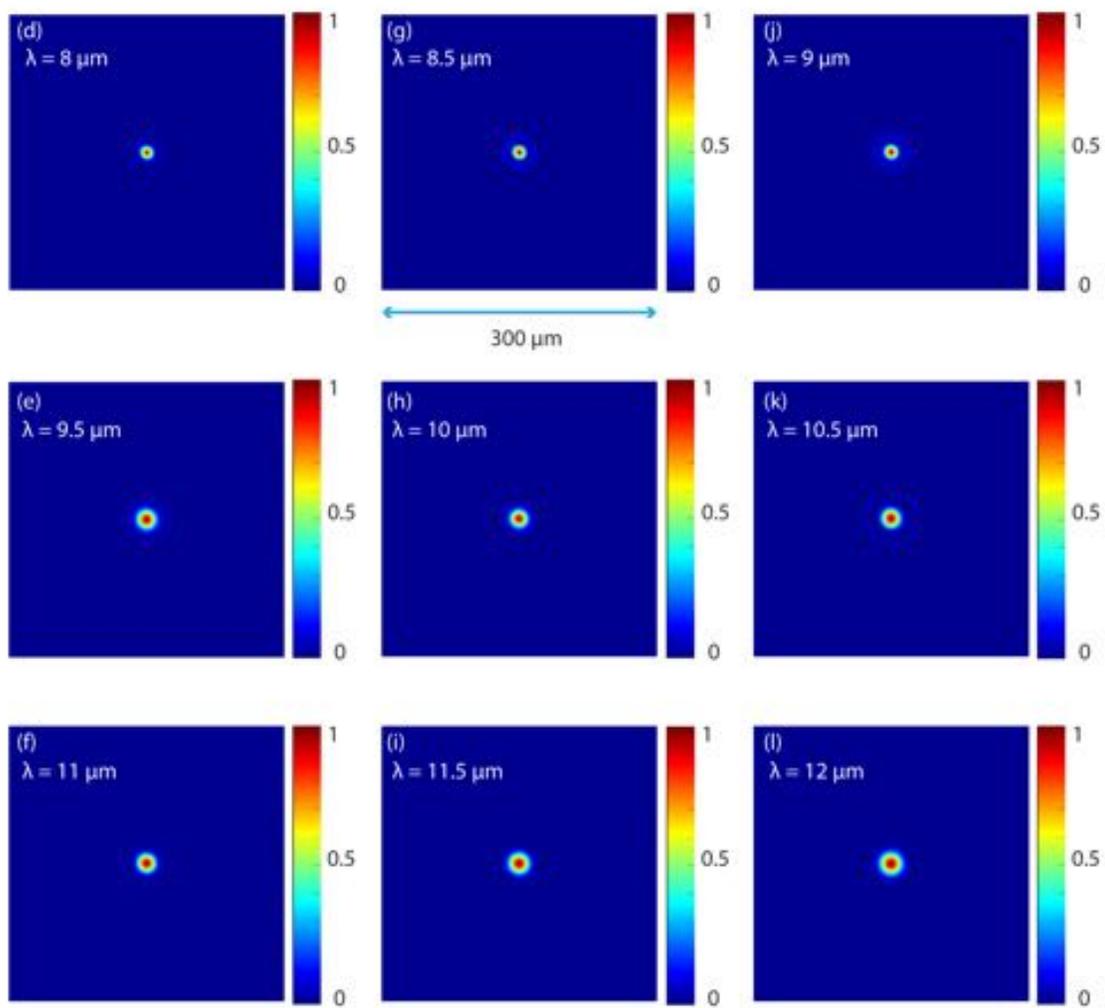

*Fig. S13:* *Simulated PSFs of the 16-level Si lens.*

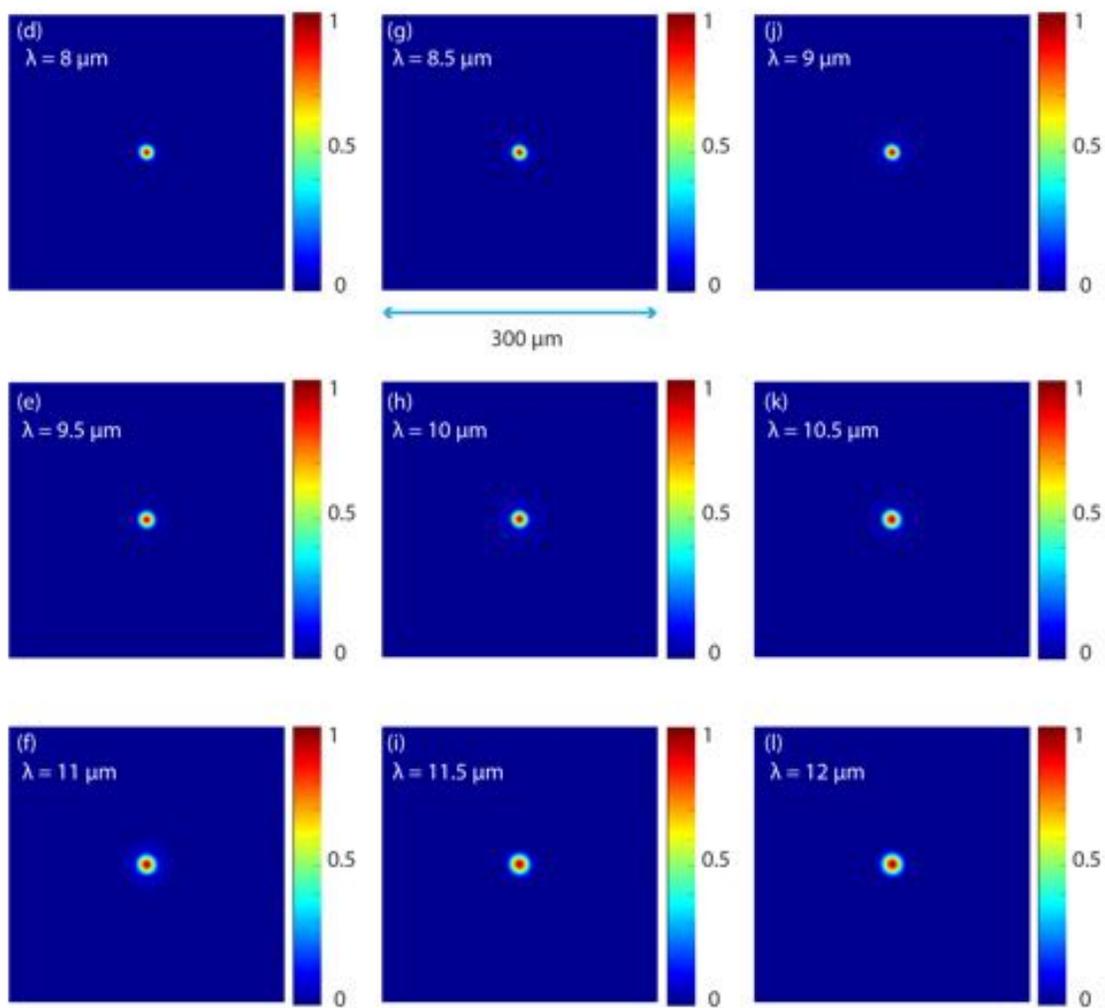

***Fig. S14:*** *Simulated PSFs of the 32-level Si lens.*

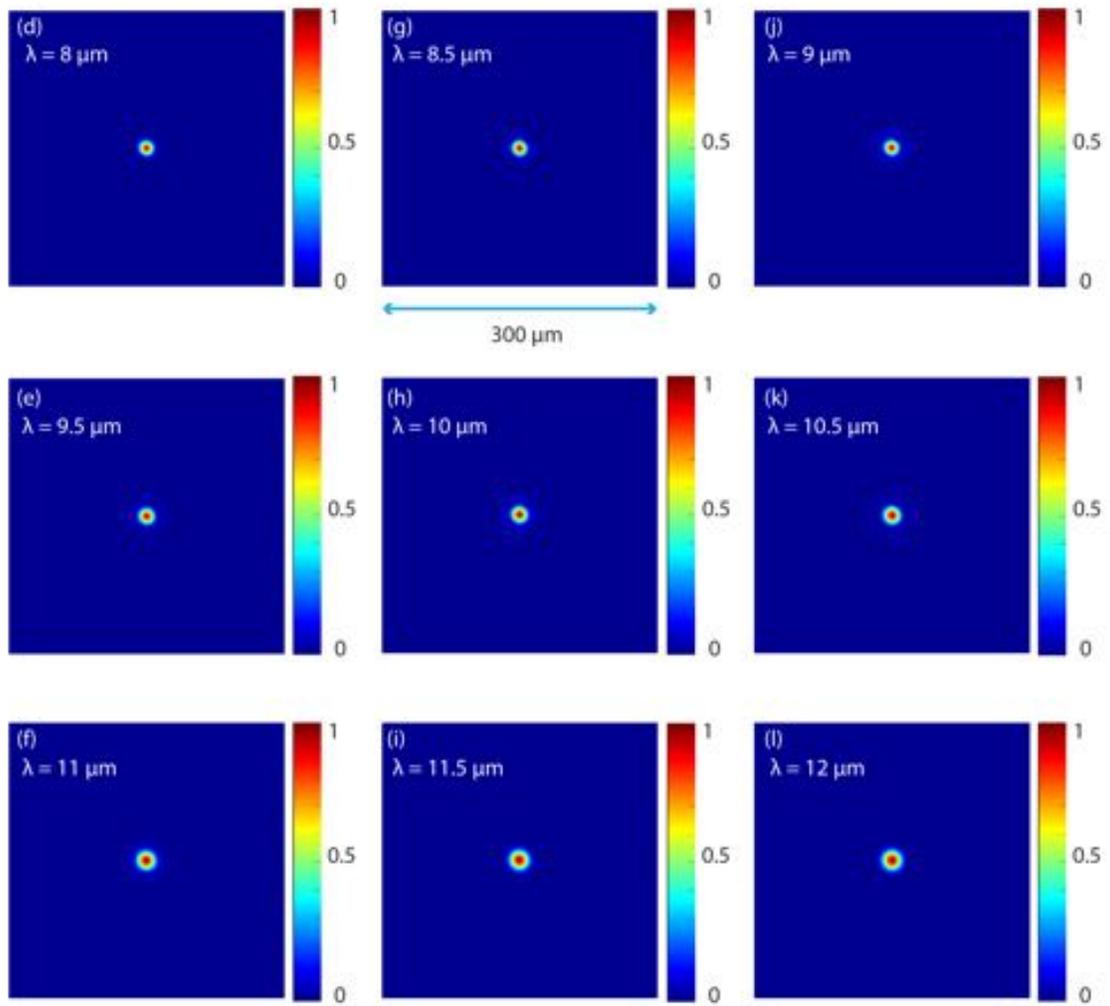

***Fig. S15:*** *Simulated PSFs of the 64-level Si lens.*